\documentclass[aps,preprint,nofootinbib,a4paper,11pt,superscriptaddress]{revtex4}% добавляется twoside для двусторонней печати

\usepackage[T1]{fontenc}
\usepackage[centertags]{amsmath}
\usepackage[mediummath]{nccmath}
\usepackage{amsfonts}
\usepackage{amssymb}
\usepackage{braket}% бра-кет обозначения
\usepackage[pdftex]{hyperref}% hypertex,pdftex
\usepackage[sort&compress]{natbib}% упорядочивает ссылки

\DeclareMathOperator{\diag}{diag}

\newcommand{\lan}{\langle}
\newcommand{\ran}{\rangle}

\begin{document}
\allowdisplaybreaks[4]% позволяет переносить многострочные формулы
\frenchspacing% уменьшение пробелов после запятых
\setlength{\unitlength}{1pt}% устанавливает единицу длины в окружении picture
%\selectlanguage{russian}

\title{Excitation of multipolar transitions in nuclei by twisted photons}

%\date{}
\author{P.~O. Kazinski}
\email[E-mail:]{kpo@phys.tsu.ru}
\affiliation{Physics Faculty, Tomsk State University, Tomsk 634050, Russia}

\author{A.~A. Sokolov}
\email[E-mail:]{alexei.sokolov.a@gmail.com}
\affiliation{Physics Faculty, Tomsk State University, Tomsk 634050, Russia}

\begin{abstract}
The explicit expression for the probability of absorption of a twisted photon by an atomic nucleus has been obtained. It is shown that photoabsorption obeys the selection rule $j\geqslant|m_\gamma|$, where $j$ is the multipolarity of the nuclear transition, when the nuclei lie near the axis along which a twisted photon with a projection of the total angular momentum $m_\gamma$ propagates. In the long-wave limit, the main contribution to the probability of absorption of a twisted photon comes from the multipole transition with $j=|m_\gamma|$. The absorption coefficient for twisted photons in a target consisting of many nuclei has been found.
\end{abstract}

\maketitle

\section{Introduction}

One of the main methods for investigating collective dynamics in atomic nuclei is the study of giant resonances excited by the interaction of nuclei with hard photons, electrons, or other charged particles \cite{IshkhKapit21}. Typically, in exciting the giant resonances, the main contribution is made by a dipole transition (the giant dipole resonance) whereas the transitions with higher multipolarity are suppressed and are poorly distinguishable against the background. In this regard, the problem arises of how to select a nuclear reaction in which the resonance under study clearly manifests itself. In the present paper, we propose to employ the twisted gamma photons for this purpose. As we will show, such photons excite the transitions of a fixed multipolarity $j$ in nuclei under certain conditions.

Twisted photons are excitations of a quantum electromagnetic field with a certain energy, a definite projection of the total angular momentum onto given axis, a fixed projection of momentum onto this axis, and a helicity \cite{Torner, Andrews, Padgett, Beams}. The projection of the total angular momentum, $m_\gamma$, can be an arbitrary integer number. Nowadays, there are sources of twisted photons with $|m_\gamma|\gtrsim10000$ in the X-ray spectrum range \cite{Fickler}. The twisted photons of MeV energies can be obtained using the inverse Compton scattering \cite{JentSerbPRL106,JentSerbEPJC71,BKL4,Ivan22,Guo23} or channeling of charged particles in crystals \cite{BogdKazTukh,AbdBogdKazTukh}. The twisted photons find applications in many areas of applied and fundamental physics. One of the reasons for the increased interest in using such photons is the possibility of controlling the rotational degrees of freedom of quantum objects by transferring a large angular momentum to them in photoabsorption processes \cite{Afanas13,Duan2019}. This idea can be employed to excite the transitions of a certain multipolarity in atomic nuclei. In this paper, we study this possibility in detail and find the explicit expression for the probability of photoabsorption of a twisted photon by a nucleus.

The paper is organized as follows. In Sec. \ref{Gen_Form_Not}, the basic notation and agreements are introduced. In Sec. \ref{Photoexc_Ampl}, the explicit expression for the amplitude of photoabsorption by the nucleus is derived. The states of the photon and of the center-of-mass of the nucleus are taken in the form of wave packets of arbitrary shapes. Section \ref{Probab_Photoexc} is devoted to the derivation of the explicit expression for the probability of photoabsorption of the twisted photon by the nucleus. In particular, in this section the selection rules governing this process are proved. In Sec. \ref{PhotoExcit_Targ}, the Bouguer formula for the absorption of a twisted photon in a target consisting of many nuclei is derived.

We use the system of units such that $\hbar=c=1$ and $e^2=4\pi\alpha$, where $\alpha$ is the fine-structure constant. The Minkowski metric has the form $\eta_{\mu\nu}=\diag(1,-1,-1,-1)$. The Greek indices run from 0 to 3 whereas the Latin indices take the integer values from 1 to 3.

\section{Basic formulas and notation}\label{Gen_Form_Not}

Define the electromagnetic field operator in the interaction picture
\begin{equation}\label{EMfield}
	\hat{\mathbf{A}}(t,\mathbf{x})=\sum_\lambda \int \frac{Vd\mathbf{k}}{(2\pi)^3} \frac{1}{\sqrt{2k_0V}}
	\Big(\hat c_\lambda(\mathbf{k}) \mathbf{e}_{\mathbf{k},\lambda}e^{-ik_{0}t+i\mathbf{k}\mathbf{x}}+\hat c^\dag_\lambda(\mathbf{k}) \mathbf{e}^*_{\mathbf{k},\lambda} e^{ik_{0}t-i\mathbf{k}\mathbf{x}}\Big),
\end{equation}
where $V$ is the normalization volume, $k_0=|\mathbf{k}|$ is the photon energy, the polarization vectors satisfy the transversality condition $(\mathbf{k},\mathbf{e}_{\mathbf{k},\lambda})=0$, and the creation-annihilation operators obey the standard commutation relations
\begin{equation}
	[\hat c_\lambda(\mathbf{k}),\hat c^\dag_{\lambda'}(\mathbf{k}')]=\frac{(2\pi)^3}{V}\delta(\mathbf{k}-\mathbf{k}')\delta_{\lambda\lambda'}, \qquad
	[\hat c_\lambda(\mathbf{k}),\hat c_{\lambda'}(\mathbf{k}')]=0.
\end{equation}
The field operator \eqref{EMfield} is represented as a superposition of plane waves with the momentum $\mathbf{k}$ and the helicity $\lambda=\pm1$.

Instead of the plane waves, the basis of mode functions used to construct the field operator can be constituted by the spherical solutions of the Maxwell equations, viz., by the multipole potentials \cite{AhiezSit}:
\begin{equation}\label{Multipoles}
	\mathbf{A}^\tau_{jm}(k_0,\mathbf{x})=\sqrt{\frac{k_0}{R_s}}\times
	\begin{cases}
	j_j(k_0r)\mathbf{Y}^j_{jm}(\theta,\varphi), & \tau = M;\\
    -\sqrt{\frac{j}{2j+1}}j_{j+1}(k_0r)\mathbf{Y}^j_{j+1,m}(\theta,\varphi)+\sqrt{\frac{j+1}{2j+1}}j_{j-1}(k_0r)\mathbf{Y}^j_{j-1,m}(\theta,\varphi), & \tau = E,
	\end{cases}
\end{equation}
where $R_s$ is the normalization radius, $j=\overline{0,\infty}$, $m=\overline{-j,j}$, and $j_n(x)$ are the spherical Bessel functions. The vector spherical harmonics satisfy the system of equations
\begin{equation}
	\begin{split}
		&\hat J^2 \mathbf{Y}^j_{lm}(\theta,\varphi)=j(j+1)\mathbf{Y}^j_{lm}(\theta,\varphi),\\
		&\hat J_3\mathbf{Y}^j_{lm}(\theta,\varphi)=m\mathbf{Y}^j_{lm}(\theta,\varphi),\\
		&\hat L^2 \mathbf{Y}^j_{lm}(\theta,\varphi)=l(l+1)\mathbf{Y}^j_{lm}(\theta,\varphi),\\
		&\hat S^2 \mathbf{Y}^j_{lm}(\theta,\varphi)=2\mathbf{Y}^j_{lm}(\theta,\varphi).
	\end{split}
\end{equation}
Thus the multipole potentials \eqref{Multipoles} are the states with the definite energy $k_0$, the square of the total angular momentum $\hat J^2$ with the eigenvalues $j(j+1)$, and its projection $\hat J_3$ with the eigenvalues $m$. In addition, the potentials of the electric and magnetic multipoles differ by parity
\begin{equation}
	\hat P \mathbf{A}^M_{jm}(k_0,\mathbf{x})=(-1)^{j+1} \mathbf{A}^M_{jm}(k_0,\mathbf{x}),\qquad \hat P \mathbf{A}^E_{jm}(k_0,\mathbf{x})=(-1)^j \mathbf{A}^E_{jm}(k_0,\mathbf{x}).
\end{equation}
The normalization condition has the form
\begin{equation}
	\int d\mathbf{x} \mathbf{A}^{\tau'*}_{j'm'}(k_0',\mathbf{x}) \mathbf{A}^\tau_{jm}(k_0,\mathbf{x}) =\frac{1}{2k_0}\frac{\pi}{R_s}\delta(k_0-k_0')\delta_{\tau\tau'}\delta_{jj'}\delta_{mm'}.
\end{equation}
In order to obtain the contributions of different multipolarities to the photoabsorption probability, we will also need the well-known representation of plane waves in term of the spherical harmonics
\begin{equation}\label{pwtumultipoles}
	\mathbf{e}_{\mathbf{k},\lambda}e^{i\mathbf{k} \mathbf{x}}=-\sqrt{\frac{R_s}{k_0}}\sum_{j=1}^\infty \sum_{m=-j}^j i^j \sqrt{2\pi(2j+1)}D^j_{m\lambda}(\varphi_k,\theta_k,0)\left\{\lambda \mathbf{A}^M_{jm}(k_0,\mathbf{x})+i\mathbf{A}^E_{jm}(k_0,\mathbf{x})\right\},
\end{equation}
where the Wigner $D$-matrix, $D^j_{mm'}(\alpha,\beta,\gamma):=e^{-im\alpha}d^j_{mm'}(\beta)e^{-im'\gamma}$, is defined as in \cite{Varshalovich}.

\section{Photoabsorption amplitude}\label{Photoexc_Ampl}

Consider the process of absorption of a photon by a nucleus
\begin{equation}
	\gamma+X\rightarrow X^*,
\end{equation}
where $X$ and $X^*$ denote the nucleus in the initial and final states, respectively. The Hamiltonian of the system under consideration has the form
\begin{equation}
	\hat H=\hat H_\gamma + \hat H_n+ \hat H_{int},
\end{equation}
where $\hat H_n$ is the Hamiltonian of nucleus, $\hat H_\gamma$ is the Hamiltonian of the free electromagnetic field. The interaction Hamiltonian in the Coulomb gauge in the interaction picture is written as
\begin{equation}\label{H_int}
	\hat H_{int}=e \int d\mathbf{x} \hat j^i(t,\mathbf{x}) \hat A_i(t,\mathbf{x})- \frac{e^2}{2}\int d \mathbf{x} d\mathbf{y} \hat j^0(t,\mathbf{x}) \Delta^{-1}(\mathbf{x}-\mathbf{y}) \hat j^0(t,\mathbf{y}),
\end{equation}
where $\Delta^{-1}(\mathbf{x}-\mathbf{y})$ is the kernel of the inverse Laplace operator, $\hat j^\mu(t,\mathbf{x})$ is the current density operator in the interaction picture. The current density operator in the interaction picture is related to the operator in the Schr\"{o}dinger picture in a standard way
\begin{equation}\label{Interac}
	\hat j^\mu(t,\mathbf{x})=e^{i\hat H_n t}\hat j^\mu(\mathbf{x}) e^{-i\hat H_n t}.
\end{equation}
The explicit form of the current density operator in the Schr\"{o}dinger picture is \cite{AhiezSit}
\begin{equation}\label{ShredTok}
	\begin{split}
		\hat j^0(\mathbf{x})&=\sum_{j=1}^A\frac{1+\tau_{zj}}{2}\delta(\mathbf{x}-\mathbf{x}_j),\\
		\hat {\mathbf{j}}(\mathbf{x})&=\sum_{j=1}^A\frac{1+\tau_{zj}}{2}\frac{\hat{\mathbf{p}}_j \delta(\mathbf{x}-\mathbf{x}_j)+\delta(\mathbf{x}-\mathbf{x}_j)\hat{ \mathbf{p}}_j}{2M}+\frac{1}{2M}\sum_{j=1}^A\left[\frac{1+\tau_{zj}}{2}\mu_p+\frac{1-\tau_{zj}}{2}\mu_n\right] \nabla\times\delta(\mathbf{x}-\mathbf{x}_j)\boldsymbol{\sigma}_j,
	\end{split}
\end{equation}
where $(1\pm\tau_{zj})/2$ are the projectors to the proton and neutron states. The plus sign corresponds to the proton state and the minus sign is for the neutron state. The magnetic moments of the proton and neutron are denoted as $\mu_p$ and $\mu_n$, $\mathbf{x}_j$ is the radius vector of the $j$-th nucleon, $A$ is the mass number, $M$ is the mass of a nucleon, $\boldsymbol{\sigma}_j$ is the spin operator. In the leading order of perturbation theory in the coupling constant $e$, the second term in the interaction Hamiltonian \eqref{H_int} can be neglected.

Let the nucleus be in the state $\ket{\mathbf{p}_i,i}$ at the instant of time  $t_1$, where $\mathbf{p}_i$ is momentum of the nucleus center-of-mass. The quantum numbers $i:=\{n_i,\mathcal{J}_i,\mathcal{M}_i\}$ describe the internal state of the nucleus, where $\mathcal{J}_i$ is the nuclear spin, $\mathcal{M}_i$ is the nuclear spin projection, and $n_i$ denote all the other quantum numbers that describe the state of the nucleus. If, on absorbing the photon prepared in the state $\ket{\mathbf{k},\lambda}$ at the instant of time $t_1$, the nucleus is found in the state $\ket{\mathbf{p}_f,f}$ at the instant of time $t_2$, then the probability amplitude of such a process is determined by the formula
\begin{equation}
	A(\mathbf{p}_f,f;\mathbf{p}_i,i,\mathbf{k},\lambda):=\braket{\mathbf{p}_f,f|\hat U^0_{t_2,0}\hat S_{t_2,t_1}\hat U^0_{0,t_1}|\mathbf{p}_i,i;\mathbf{k},\lambda},
\end{equation}
where the operator $\hat U^0_{t_2,t_1}$ describes a free evolution and the $\hat{S}$-operator has the form
\begin{equation}
	\hat S_{t_2,t_1}=1-ie\int d^4x\hat j^i(t,\mathbf{x}) \hat A_i(t,\mathbf{x}),
\end{equation}
in the leading nontrivial order of the perturbation theory. Using the standard relations for the creation-annihilation operators,
\begin{equation}
	\hat U^0_{0,t}\hat c_\alpha \hat U^0_{t,0}=e^{-iE_\alpha t}\hat c_\alpha,\qquad U^0_{t_2,t_1}\ket{0}=e^{-iE_0(t_2-t_1)}\ket{0},
\end{equation}
where $E_\alpha$ is the energy of the one-particle state $\alpha$ and $E_0$ is the vacuum energy, the scattering amplitude becomes
\begin{equation}\label{a0}
	A(\mathbf{p}_f,f;\mathbf{p}_i,i,\mathbf{k},\lambda)=-iee^{-iE_0(t_2-t_1)}e^{-iE_f t_2}e^{i(E_i+k_0)t_1}\int_{t_1}^{t_2}d^4x
	\braket{\mathbf{p}_f,f|\hat j^i(t,\mathbf{x})|\mathbf{p}_i,i}\braket{0|\hat A_i(t,\mathbf{x})|\mathbf{k},\lambda}.
\end{equation}

If the photon and the nucleus are prepared in the states described by two arbitrary wave packets at the instant of time $t_1$,
\begin{equation}
	\ket{\tilde\psi}:=\sqrt{\frac{V}{(2\pi)^3}}\sum_\lambda \int d\mathbf{k} \tilde\psi(\mathbf{k},\lambda)\ket{\mathbf{k},\lambda},\qquad\ket{\tilde\phi}:=\sqrt{\frac{V}{(2\pi)^3}}\int d\mathbf{p}_i \tilde\phi(\mathbf{p}_i)\ket{\mathbf{p}_i,i},
\end{equation}
and these wave functions are normalized by the condition
\begin{equation}\label{norm}
	\sum_\lambda \int d\mathbf{k}|\tilde\psi(\mathbf{k},\lambda)|^2=1,\qquad\int d\mathbf{p}_i|\tilde\phi(\mathbf{p}_i)|^2=1,	
\end{equation}
then the scattering amplitude of such a process is obtained with the aid of the amplitude \eqref{a0} as
\begin{equation}\label{amplt_1}
	\begin{split}
		A(\mathbf{p}_f,f;\tilde\phi,i,\tilde\psi):=&-ie\frac{V}{(2\pi^3)}e^{-iE_0(t_2-t_1)}e^{-iE_f t_2}\sum_\lambda
		\int d\mathbf{k} e^{ik_0t_1}\tilde \psi(\mathbf{k},\lambda)\int d\mathbf{p}_i e^{iE_it_1}\tilde \phi(\mathbf{p}_i)\times\\
		&\times\int_{t_1}^{t_2}d^4x \braket{\mathbf{p}_f,f|\hat j^i(x)|\mathbf{p}_i,i}\braket{0|\hat A_i(x)|\mathbf{k},\lambda}.
	\end{split}
\end{equation}
It is convenient to specify the form of the wave functions of the photon and the nucleus at the instant of time $t=0$,
\begin{equation}\label{wtwp}
	\psi(\mathbf{k},\lambda):=e^{ik_0t_1}\tilde \psi(\mathbf{k},\lambda),\qquad\phi(\mathbf{p}_i)=e^{iE_it_1}\tilde \phi(\mathbf{p}_i),
\end{equation}
assuming that they evolve freely from the moment $t=t_1$ to the moment $t=0$. Then the functions $\tilde \psi(\mathbf{k},\lambda)$, $\tilde \phi(\mathbf{p}_i)$ are found from equalities \eqref{wtwp}. Notice that the functions $\psi(\mathbf{k},\lambda)$, $\phi(\mathbf{p}_i)$ are normalized by the same condition \eqref{norm}.

Putting $t_1\rightarrow -\infty$, $t_2\rightarrow \infty$ in \eqref{amplt_1} and discarding the irrelevant phase factors, we come to
\begin{equation}
	A(\mathbf{p}_f,f;\phi,i,\psi)=\frac{eV}{(2\pi)^3}\sum_\lambda \int d\mathbf{k} d\mathbf{p}_i \psi(\mathbf{k},\lambda)\phi(\mathbf{p}_i)\int d^4x\braket{\mathbf{p}_f,f|
		\hat j^i(x)|\mathbf{p}_i,i}\braket{0|\hat A_i(x)|\mathbf{k},\lambda}.
\end{equation}
Now we substitute the matrix element of the electromagnetic field,
\begin{equation}
	\braket{0|\hat{\mathbf{A}}(x)|\mathbf{k},\lambda}= \frac{1}{\sqrt{2k_0V}}\mathbf{e}_{\mathbf{k},\lambda}e^{-ik_0t+i\mathbf{k}\mathbf{x}},
\end{equation}
pass to the Schr\"{o}dinger picture \eqref{Interac}, and take the integral over time $t$:
\begin{equation}\label{a1}
	A(\mathbf{p}_f,f;\phi,i,\psi)=\frac{eV}{(2\pi)^2}\sum_\lambda \int \frac{d\mathbf{k} d\mathbf{p}_i}{\sqrt{2k_0 V}}
	\psi(\mathbf{k},\lambda)\phi(\mathbf{p}_i)\delta(E_f-E_i-k_0)\int d\mathbf{x} \braket{\mathbf{p}_f,f|\hat j_i(\mathbf{x})|\mathbf{p}_i,i}e^i_{\mathbf{k},\lambda}e^{i\mathbf{k}\mathbf{x}}.
\end{equation}
In order to evaluate the matrix element of the current density operator \eqref{ShredTok}, it is convenient to introduce the relative coordinates and momenta by making the replacement
\begin{equation}
	\mathbf{x}_j'=\mathbf{x}_j-\mathbf{R},\qquad \hat{\mathbf{p}}_j'=\hat{\mathbf{p}}_j-\frac{1}{A}\hat{\mathbf{p}},
\end{equation}
where $\mathbf{R}$ and $\hat{\mathbf{p}}$ are the radius vector and the momentum of the center-of-mass of the nucleus. Having replaced the coordinates, the state of the nucleus is factorized
\begin{equation}
	\ket{\mathbf{p}_i,i}=\frac{1}{\sqrt{V}}e^{i\mathbf{p}_i \mathbf{R}}\ket{i},\qquad \ket{\mathbf{p}_f,f}=\frac{1}{\sqrt{V}}e^{i\mathbf{p}_f \mathbf{R}}\ket{f}.
\end{equation}
Then we shift the integration variable $\mathbf{x}'=\mathbf{x}-\mathbf{R}$ and make the replacement in the current density operator
\begin{equation}
	\hat {\mathbf{j}}(\mathbf{x})\rightarrow \hat {\mathbf{j}}(\mathbf{x}')+\frac{\hat {\mathbf{p}} \hat j^0(\mathbf{x}')+\hat j^0(\mathbf{x}') \hat {\mathbf{p}} }{2AM}.
\end{equation}
Then the amplitude \eqref{a1} is written as
\begin{equation}\label{a2}
	\begin{split}
		A(\mathbf{p}_f,f;\phi,\psi)=&\,2\pi e\sum_\lambda \int \frac{d\mathbf{k} d\mathbf{p}_i}{\sqrt{2k_0 V}}
		\psi(\mathbf{k},\lambda)\phi(\mathbf{p}_i)\delta(E_f-E_i-k_0)\delta(\mathbf{p}_f-\mathbf{p}_i-\mathbf{k})\times\\
		&\times\Big[\int d\mathbf{x}' \braket{f|\hat j_i(\mathbf{x}')|i}e^i_{\mathbf{k},\lambda}e^{i\mathbf{k}\mathbf{x}'}+\int d\mathbf{x}'		 \braket{f|j^0(\mathbf{x}')|i}\frac{\left(\mathbf{p}_f+\mathbf{p}_i,\mathbf{e}_{\mathbf{k},\lambda}\right)}{2AM}e^{i\mathbf{k}\mathbf{x}'}\Big].
	\end{split}
\end{equation}
Simplifying the second term,
\begin{equation}
	\braket{f|j^0(\mathbf{x}')|i} \frac{\left(\mathbf{p}_f+\mathbf{p}_i,\mathbf{e}_{\mathbf{k},\lambda}\right)}{2AM} =
	\braket{f|j^0(\mathbf{x}')|i}\frac{\left(\mathbf{p}_i,\mathbf{e}_{\mathbf{k},\lambda}\right)}{AM},
\end{equation}
where we have taken into account the transversality condition $(\mathbf{k},\mathbf{e}_{\mathbf{k},\lambda})=0$, it becomes evident that this term describes the transition current of the nucleus as a whole. The first term in the amplitude \eqref{a2} describes the transition current between the internal states of the nucleus. This matrix element can only be found by choosing a specific model for the interaction of the components of the nucleus. If the standard deviation of momenta in the wave packet of the center-of-mass of the nucleus, $\sigma$, is much less than the average momentum of nucleons in the nucleus, $\lan p_n\ran$,
\begin{equation}\label{usl0}
	\frac{\sigma}{A \lan p_n\ran}\ll1,
\end{equation}
then the second term in the amplitude \eqref{a2} can be neglected. Recall that for Gaussian wave packets $1/\sigma$ is of order of the size of the wave packet of the nucleus center-of-mass in the coordinate space. Therefore, the condition \eqref{usl0} is usually satisfied.

Substituting the expansion of the plane wave in the form \eqref{pwtumultipoles} into the amplitude \eqref{a2}, we obtain
\begin{equation}\label{a3}
	A(\mathbf{p}_f,f;\phi,\psi)=2\pi e\sqrt{\frac{R_s}{2V}} \sum_\lambda \int\frac{d\mathbf{k}}{k_0}d\mathbf{p}_i
	\psi(\mathbf{k},\lambda)\phi(\mathbf{p}_i)\delta(E_f-E_i-k_0)\delta(\mathbf{p}_f-\mathbf{p}_i-\mathbf{k})f_\lambda(\mathbf{k}),
\end{equation}
where the function,
\begin{equation}
	f_\lambda(\mathbf{k}):=\sum_{j=1}^\infty \sum_{m=-j}^j i^j \sqrt{2\pi (2j+1)}D^j_{m\lambda}(\varphi_k,\theta_k,0)
	\left\{\lambda M^M_{jm}(k_0)+i M^E_{jm}(k_0)\right\},
\end{equation}
has been defined to shorten the notation, and the standard notation has been introduced for the matrix elements of nuclear multipole operators
\begin{equation}\label{M_matr}
	M^\tau_{jm}(k_0):=\int d\mathbf{x}' \braket{f|\mathbf{j}(\mathbf{x}')|i}\mathbf{A}^\tau_{jm}(k_0,\mathbf{x}').
\end{equation}
Let us simplify the amplitude \eqref{a3}. To this end, we rewrite the argument of the first delta function introducing explicitly the excitation energy of the nucleus $\varepsilon$:
\begin{equation}
	E_f-E_i-k_0=\frac{\mathbf{p}^2_f}{2AM}+\varepsilon-\frac{\mathbf{p}_i^2}{2AM}-k_0=0.
\end{equation}
Then
\begin{equation}
	k_0=\varepsilon-\frac{(\mathbf{p}_i,\mathbf{k})}{AM}-\frac{\mathbf{k}^2}{2AM}.
\end{equation}
The term $(\mathbf{p}_i,\mathbf{k})/(AM)$ describes the Doppler effect due to motion of the center-of-mass of the nucleus. The term $\mathbf{k}^2/(2AM)$ is responsible for the quantum recoil.

The Doppler effect can be neglected provided the standard deviation of momenta in the wave packet of the center-of-mass of the nucleus is much less than the mass of the nucleus
\begin{equation}\label{usl1}
	\frac{\sigma}{AM}\ll1.
\end{equation}
The term responsible for the quantum recoil can be discarded if the energy of the photon is much less then the mass of the nucleus
\begin{equation}\label{usl2}
	\frac{k_0}{AM}\ll1.
\end{equation}
Therefore, we suppose that $k_0\approx\varepsilon $. Then we can evaluate the integral over $k_3$ in \eqref{a3} taking into account that
\begin{equation}
	\frac{\delta(\varepsilon-k_0)}{k_0}=\frac{\delta(k_3-\tilde k_3(k_\perp))}{\tilde k_3},\qquad \tilde k_3(k_\perp):=\sqrt{\varepsilon^2-k_\perp^2}.
\end{equation}
The delta function expressing the momentum conservation law in \eqref{a3} allows one to perform integration with respect to the momentum $\mathbf{p}_i$. As a result, we obtain
\begin{equation}\label{a4}
	A(\mathbf{p}_f,f;\phi,i,\psi)=2\pi e\sqrt{\frac{R_s}{2V}} \sum_\lambda \int\frac{d\mathbf{k}_\perp}{k_3}
	\psi(\mathbf{k},\lambda)\phi(\mathbf{p}_f-\mathbf{k})f_\lambda(\mathbf{k})|_{k_3=\tilde k_3(k_\perp)}.
\end{equation}

\section{Probability of photoabsorption}\label{Probab_Photoexc}

The probability of photoabsorption with the transition of the nucleus from the internal state $i$ to the internal state $f$ reads as
\begin{equation}\label{IncProb1}
	P(i\rightarrow f)=\int \frac{Vd\mathbf{p}_f}{(2\pi)^3}\left|A(\mathbf{p}_f,f;\phi,i,\psi)\right|^2.
\end{equation}
The amplitude \eqref{a4} has been obtained in a general form for the arbitrary wave packets satisfying the conditions \eqref{usl0}, \eqref{usl1}, and \eqref{usl2}. To calculate explicitly the remaining integrals in the amplitude \eqref{a4}, it is necessary to specify the shapes of the wave packets. We take the twisted state of the photon,
\begin{equation}
	\psi_{k^0_\perp k^0_3m_\gamma\lambda_0}(\mathbf{k},\lambda)=
	C_\gamma k^{|m_\gamma|}_\perp e^{-\frac{[k_\perp^2-(k_\perp^0)^2]^2}{4\sigma^4_\perp}} e^{-\frac{(k_3-k_3^0)^2}{4\sigma^2_3}} e^{im_\gamma\varphi_k}\delta_{\lambda\lambda_0},
\end{equation}
with a certain projection of the total angular momentum $m_\gamma$ and a helicity $\lambda_0$ \cite{KazRyakExcit}. In the limit $\sigma_\perp\rightarrow0$, $\sigma_3\rightarrow0$, this state goes into the standard Bessel state \cite{JaurHac,BiaBirBiaBir,PRA97}. As for the center-of-mass of the nucleus, we choose the Gaussian wave packet
\begin{equation}
	\phi(\mathbf{p})=C_n e^{-\frac{\mathbf{p}^2}{4\sigma^2}}e^{-i\mathbf{p}\mathbf{b}_\perp}.
\end{equation}
The normalization constants $C_\gamma$, $C_n$ are determined by the normalization conditions \eqref{norm}. The vector $\mathbf{b}_\perp:=\{b_x,b_y,0\}$ is the impact parameter between the axis along which the twisted photon propagates and the center-of-mass of the nucleus.

The photoabsorption probability \eqref{IncProb1} involves the expression
\begin{equation}
	\phi(\mathbf{p}_f-\mathbf{k})\phi^*(\mathbf{p}_f-\mathbf{k}')=C_n^2 e^{-\frac{(\mathbf{p}_f-\mathbf{k})^2+(\mathbf{p}_f-\mathbf{k}')^2}{4\sigma^2}}e^{-i\Delta\mathbf{k}\mathbf{b}_\perp},
\end{equation}
where $\Delta\mathbf{k}:=\mathbf{k}'-\mathbf{k}$. Completing the square in the exponent, it can be integrated over the momentum $\mathbf{p}_f$:
\begin{equation}
	\int d\mathbf{p}_f \phi(\mathbf{p}_f-\mathbf{k})\phi^*(\mathbf{p}_f-\mathbf{k}')=
	\int d\mathbf{p}_f C_n^2e^{-\frac{(\mathbf{p}_f+\frac{\Delta\mathbf{k}}{2})^2}{2\sigma^2}}e^{-\frac{\Delta\mathbf{k}^2}{8\sigma^2}}e^{-i\Delta\mathbf{k}\mathbf{b}} =e^{-\frac{\Delta\mathbf{k}^2}{8\sigma^2}}e^{-i\Delta\mathbf{k}\mathbf{b}_\perp}.
\end{equation}
Then the photoabsorption probability \eqref{IncProb1} takes the form
\begin{equation}
	P(i\rightarrow f)=\alpha R_s \sum_{\lambda,\lambda'}
	\int \frac{d\mathbf{k}_\perp}{k_3}\frac{d\mathbf{k}_\perp'}{k_3'}\psi(\mathbf{k},\lambda)\psi^*(\mathbf{k}',\lambda') e^{-\frac{\Delta\mathbf{k}^2}{8\sigma^2}} e^{-i\Delta\mathbf{k}\mathbf{b}_\perp}f_\lambda(\mathbf{k})f_{\lambda'}(\mathbf{k}')\Big|_{
		\substack{k_3=\tilde k_3(k_\perp) \\k_3'=\tilde k_3(k_\perp')}}.
\end{equation}
We perform the integrals over $k_\perp$ using the perturbation theory with respect to the small parameter $\sigma_\perp$. In this case, to simplify the resulting expression, we assume that
\begin{equation}
	k_3^0=\tilde k_3(k_\perp^0)=\sqrt{\varepsilon^2-(k_\perp^0)^2}.
\end{equation}
For the perturbation theory with respect to $\sigma_\perp$ to be applicable, the following conditions have to be met
\begin{equation}\label{conds}
	\frac{k_\perp^0\sigma_\perp}{(k_3^0)^2}\ll1,\quad \frac{(2|m_\gamma|+1)\sigma_\perp}{k_\perp^0}\ll1,\quad
	\frac{k_\perp^0\sigma_\perp}{\sigma^2}\ll1,\quad  \frac{\sigma_3}{k_3^0}\frac{k_\perp^0\sigma_\perp}{\sigma^2}\ll1,\quad  \sigma_\perp b_\perp\ll1, \quad \frac{\left(\mathcal{J}_f+\mathcal J_i+1\right)\sigma_\perp}{k_\perp^0}\ll1.
\end{equation}
Then, evaluating the integrals over $k_\perp$, $k'_\perp$, we obtain in the leading order
\begin{equation}
	P(i\rightarrow f)=\pi \alpha R_s |C_\gamma|^2 \sigma_\perp^4 \frac{(k_\perp^0)^{2|m_\gamma|}}{(k_3^0)^2}\int_0^{2\pi} d \varphi_k d \varphi_k' e^{im_\gamma(\varphi_k-\varphi_k')}e^{-\frac{\Delta\mathbf{k}_\perp^2}{8\sigma^2}}e^{-i\Delta \mathbf{k}_\perp \mathbf{b}_\perp}f_{\lambda_0}(\mathbf{k})f^*_{\lambda_0}(\mathbf{k}'),
\end{equation}
where it is assumed that $k_3=k_3'=k_3^0$, $k_\perp=k_\perp'=k_\perp^0$. The normalization constant equals to
\begin{equation}
	|C_\gamma|^2=\frac{1}{2\pi^2 \sigma_3 \sigma_\perp^2 (k_\perp^0)^{2|m_\gamma|}}+O(\sigma_\perp^2),
\end{equation}
in the leading order with respect to $\sigma_\perp$.

Taking into account that
\begin{equation}
	\begin{split}
    \Delta\mathbf{k}_\perp^2&=\left(\mathbf{k}_\perp'-\mathbf{k}_\perp\right)^2=2(k_\perp^0)^2-2(k_\perp^0)^2\cos{(\varphi_k'-\varphi_k)},\\
		\Delta\mathbf{k}_\perp \mathbf{b}&=k_\perp^0 b_\perp\left(\cos{(\varphi_k'-\varphi_b)}-\cos{(\varphi_k-\varphi_b)}\right),
	\end{split}
\end{equation}
the remaining integrals containing the angles $\varphi_k$, $\varphi_k'$ are written as
\begin{equation}
	\int_0^{2\pi}d\varphi_k d\varphi_k'   e^{ik_\perp^0b_\perp
		\cos{(\varphi_k-\varphi_b)}-i(m-m_\gamma)\varphi_k} e^{\frac{(k_\perp^0)^2}{4\sigma^2}\cos{(\varphi_k-\varphi_k')}} \ e^{-ik_\perp^0b_\perp\cos{(\varphi_k'-\varphi_b)}+i(m'-m_\gamma)\varphi_k'}.
\end{equation}
In order to reduce this integral to the product of two independent integrals, we employ the Jacobi-Anger expansion,
\begin{equation}
	e^{z\cos{\varphi}}=\sum_{n=-\infty}^\infty e^{i n \varphi}I_n(z),
\end{equation}
where $I_n(z)$ is the Bessel function of imaginary argument of order $n$. Then, using the integral representation of the Bessel function,
\begin{equation}
	J_m(z)=i^{-m}\int^{2\pi}_0 \frac{d\varphi}{2\pi}e^{-im\varphi+i z \cos{\varphi}},\quad m \in \mathbb Z,
\end{equation}
we eventually obtain
\begin{equation}\label{prob_photo1}
	\begin{split}
		P(i\rightarrow f)=&\,4\pi^2
		\frac{\alpha R_s\sigma_\perp^2}{\sigma_3(k_3^0)^2} e^{-\frac{(k_\perp^0)^2}{4\sigma^2}} \sum_{j,j'}\sum_{m,m'}  i^{j+m-j'-m'} e^{i(m'-m)\varphi_b} \sqrt{(2j+1)(2j'+1)} d^j_{m\lambda_0}(\theta_k^0)d^{j'}_{m'\lambda_0}(\theta_k^0)\times \\
		&\times \sum_{n=-\infty}^\infty I_n\Big(\frac{(k_\perp^0)^2}{4\sigma^2}\Big)J_{m-m_\gamma-n}(k_\perp^0b_\perp)J_{m'-m_\gamma-n}(k_\perp^0b_\perp)\times\\
		&\times\left\{\lambda_0 M^M_{jm}(\varepsilon)+i M^E_{jm}(\varepsilon)\right\} \left\{\lambda_0 M^{M*}_{j'm'}(\varepsilon)-i M^{E*}_{j'm'}(\varepsilon)\right\},
	\end{split}
\end{equation}
where $\theta^0_k:=\arctan (k^0_\perp/k^0_3)$. Notice that this expression does not depend on the normalization radius $R_s$ since the matrix elements of the multipole transitions \eqref{M_matr} contain the factor $1/\sqrt{R_s}$.

Usually, the initial state of the nucleus is mixed with respect to the spin projections $\mathcal M_i$ and the states with different $\mathcal M_i$ are realized with equal probability. The projection of the spin of the final state of the nucleus, $\mathcal M_f$, is also not recorded, as a rule. Therefore, we consider the probability of transition from the state with spin $\mathcal J_i$ to the state with spin $\mathcal J_f$, sum the photoabsorption probability \eqref{prob_photo1} over the projections of the final spin $\mathcal M_f$, and average over the initial ones $\mathcal M_i$:
\begin{equation}\label{prob_photo2}
	P(\mathcal{J}_i\rightarrow \mathcal{J}_f):=\frac{1}{2\mathcal J_i+1}\sum_{\mathcal M_i=-\mathcal J_i}^{\mathcal J_i}
	\sum_{\mathcal{M}_f=-\mathcal{J}_f}^{\mathcal{J}_f} P(i\rightarrow f).
\end{equation}
The sums over nuclear spin projections can be calculated explicitly. Indeed, as long as the matrices $M^\tau_{jm}(\varepsilon)$ are irreducible tensors, the Wigner-Eckart theorem applies to them
\begin{equation}\label{Wign_Eck}
	M^\tau_{jm}(\varepsilon)=\frac{1}{\sqrt{2\mathcal{J}_f+1}}C^{\mathcal{J}_f\mathcal{M}_f}_{\mathcal{J}_i\mathcal{M}_i jm}M^\tau_{j}(\varepsilon),
\end{equation}
where $C^{j_3m_3}_{j_1m_1j_2m_2}$ are the Clebsch-Gordan coefficients and $M^\tau_{j}(\varepsilon):=\braket{n_f,\mathcal{J}_f||\hat M^\tau_{j}(\varepsilon)||n_i,\mathcal{J}_i}$ are the reduced matrix elements independent of the spin projections. Substituting \eqref{Wign_Eck} into \eqref{prob_photo1}, \eqref{prob_photo2} and using the property of the Clebsch-Gordan coefficients,
\begin{equation}	 \sum_{\mathcal{M}_i=-\mathcal{J}_i}^{\mathcal{J}_i}\sum_{\mathcal{M}_f=-\mathcal{J}_f}^{\mathcal{J}_f}C^{\mathcal{J}_f\mathcal{M}_f}_{\mathcal{J}_i\mathcal{M}_i jm} C^{\mathcal{J}_f\mathcal{M}_f}_{\mathcal{J}_i\mathcal{M}_i j'm'}=\frac{2\mathcal{J}_f+1}{2j+1}\delta_{jj'}\delta_{mm'},
\end{equation}
we deduce
\begin{equation}	 \frac{1}{2\mathcal{J}_i+1}\sum_{\mathcal{M}_i=-\mathcal{J}_i}^{\mathcal{J}_i}\sum_{\mathcal{M}_f=-\mathcal{J}_f}^{\mathcal{J}_f}\left\{\lambda_0 M^M_{jm}+iM^E_{jm}\right\}\left\{\lambda_0 M^{M*}_{jm}-iM^{E*}_{jm}\right\}=\frac{\delta_{jj'}\delta_{mm'}}{(2\mathcal{J}_i+1)(2j+1)}|\lambda_0 M^M_{j}+iM^E_{j}|^2.
\end{equation}
The product of the nucleus parities in the initial and final states $\pi_i$ and $\pi_f$ must be equal to the parity $\pi^\tau_j$ of the multipole operator of the type $\tau$. The parities of the electric and magnetic operators for a fixed $j$ are different. Hence
the product $M^M_j(\varepsilon)M^E_{j}(\varepsilon)$ equals to zero. Then we expand the modulus squared in the last expression and obtain
\begin{equation}\label{PhotProbBi}
	P(\mathcal{J}_i\rightarrow \mathcal{J}_f)=\frac{4\pi^2 \alpha R_s}{2\mathcal{J}_i+1}
	\frac{\sigma_\perp^2 e^{-\frac{(k_\perp^0)^2}{4\sigma^2}}}{\sigma_3(k_3^0)^2} \sum_{j=1}^\infty\sum_{m=-j}^j \sum_{n=-\infty}^\infty
	I_n\Big(\frac{(k_\perp^0)^2}{4\sigma^2}\Big) J^2_{m-m_\gamma-n}(k_\perp^0b_\perp) \Big[d^j_{m\lambda_0}(\theta_k^0)\Big]^2 \sum_{\tau=E,M}|M^\tau_{j}(\varepsilon)|^2.
\end{equation}

If the center-of mass of the nucleus is close to the axis of propagation of the twisted photon and the following conditions are met,
\begin{equation}\label{Assympt}
	k_0 b_\perp \ll1,\qquad \frac{k_0}{\sigma}\ll1,
\end{equation}
then the Bessel functions can be replaced by the Kronecker deltas
\begin{equation}
	I_n(0)=\delta_{n0},\qquad J_{m-m_\gamma-n}(0)=\delta_{m-m_\gamma,n}.
\end{equation}
For the twisted photon with the energy of order $k_0\approx10$ keV the conditions \eqref{Assympt} are satisfied when
\begin{equation}
	b_\perp\ll 1\;\text{nm}, \qquad \sigma\gg 1\;\text{keV}.
\end{equation}
Such parameters can be achieved, for example, by using the Paul traps \cite{PRL129} to confine the nuclei and the source of hard twisted photons created by the inverse Compton scattering \cite{JentSerbPRL106,JentSerbEPJC71,BKL4,Ivan22,Guo23} or by channeling \cite{BogdKazTukh,AbdBogdKazTukh}. Then the probability of photoabsorption turns into
\begin{equation}\label{prob_photo3}
	P(\mathcal{J}_i\rightarrow \mathcal{J}_f)=\frac{4\pi^2 \alpha R_s}{2\mathcal{J}_i+1}  \frac{\sigma_\perp^2}{\sigma_3(k_3^0)^2}\sum_{j\geqslant \max{(|m_\gamma|,1)}} \Big[d^j_{m_\gamma\lambda_0}(\theta_k^0)\Big]^2\sum_{\tau=E,M}|M^\tau_{j}(\varepsilon)|^2.
\end{equation}
It is clear from this formula that, in the case $|m_\gamma|>1$, only the multipole transitions with $j\geqslant|m_\gamma|$ contribute to the photoabsorption probability. The contributions of lower multipolarity appear in the next orders of expansion in $k_\perp^0b_\perp$ and are strongly suppressed. Moreover, if the long-wave approximation is justified, the leading contribution comes from the terms with the minimum possible multipolarity $j=|m_\gamma|$ provided they are not prohibited by other selection rules.

For $m_\gamma=\lambda_0$ and $\theta_k^0\rightarrow0$, formula \eqref{prob_photo3} reproduces the well-known result for photoabsorption of a plane wave photon \cite{AhiezSit} generalized to the case when the initial state of the photon is given by a Gaussian wave packet. In this case,
\begin{equation}\label{prob_photo3_pl}
	P(\mathcal{J}_i\rightarrow \mathcal{J}_f)=\frac{4\pi^2 \alpha R_s}{2\mathcal{J}_i+1}  \frac{\sigma_\perp^2}{\sigma_3(k_3^0)^2}
	\sum_{j\geqslant1} \sum_{\tau=E,M}|M^\tau_{j}(\varepsilon)|^2.
\end{equation}
Comparing \eqref{prob_photo3} with \eqref{prob_photo3_pl}, we see that the ratio of the probability of excitation of the $j$-th multipole transition by a twisted photon to the probability of excitation of the same transition by a plane wave photon is equal to
\begin{equation}
	\Big[d^j_{m_\gamma\lambda_0}(\theta_k^0)\Big]^2\leqslant1.
\end{equation}
Nevertheless, in photoabsorption of a twisted photon with $|m_\gamma|>1$, the multipole transition with $j=|m_\gamma|$ is not overlapped by transitions of lower multipolarities and can be studied separately.

\section{Photoabsorption in a target consisting of many nuclei}\label{PhotoExcit_Targ}

Let us obtain the probability of photoabsorption of a twisted photon by a target consisting of many nuclei. Let $P_k$ be the probability of photoabsorption by one nucleus specified by formula \eqref{PhotProbBi} with the impact parameter $\mathbf{b}^k_\perp$. We assume that initially all the nuclei are in the ground state. Then the probability of photoexcitation of at least one nucleus from the state with quantum numbers $n_i,\mathcal{J}_i$ to the state with quantum numbers $n_f,\mathcal{J}_f$ reads
\begin{equation}
	P_T(\mathcal{J}_i\rightarrow\mathcal{J}_f)=1-\prod_{k=1}^N(1-P_k)=1-\exp{\Big(\sum_{k=1}^N \ln{(1-P_k)}\Big)},
\end{equation}
where $N$ is the number of nuclei in the target. If the probability of every event of photoabsorption by the $k$-th nucleus is small, $P_k\ll1$, for any $k$, then we deduce in the leading order
\begin{equation}
	P_T(\mathcal{J}_i\rightarrow\mathcal{J}_f)=1-\exp{\Big(-\sum_{k=1}^N P_k\Big)}.
\end{equation}
Instead of calculating such sums, it is convenient to take a continuous distribution of nuclei in the target with the density distribution $\rho(\mathbf{b}_\perp)$. The function $\rho(\mathbf{b}_\perp)$ is normalized by the condition
\begin{equation}
	N= \int d\mathbf{b}_\perp \rho(\mathbf{b}_\perp).
\end{equation}
Then the probability of photoabsorption by the target is given by
\begin{equation}
	P_T(\mathcal{J}_i\rightarrow\mathcal{J}_f)=1-\exp{\Big(-\int d\mathbf{b}_\perp \rho(\mathbf{b}_\perp) P(\mathbf{b}_\perp)\Big)},
\end{equation}
where $P(\mathbf{b}_\perp)$ is defined by formula \eqref{PhotProbBi}.

Consider the particular case when the distribution of nuclei in the target with respect to the impact parameter is given by the Gaussian function
\begin{equation}
	\rho(\mathbf{b}_\perp)=\frac{N}{2\pi w^2}e^{-\frac{\mathbf{b}_\perp^2}{2w^2}}.
\end{equation}
The dependence of \eqref{PhotProbBi} on the impact parameter is contained solely in the square of the Bessel function. Therefore, the integral over $\mathbf{b}_\perp$ is readily evaluated \cite{Prud2}:
\begin{equation}
	\frac{N}{2\pi w^2}\int d\mathbf{b}_\perp e^{-\frac{b_\perp^2}{2w^2}} J^2_{m-m_\gamma-n}(k_\perp^0 b_\perp)=N e^{-(k_\perp^0 w)^2}I_{m-m_\gamma-n}\big((k_\perp^0 w)^2\big).
\end{equation}
The sum over $n$ appearing in \eqref{PhotProbBi} is nothing but the addition theorem
\begin{equation}
	\sum_{n=-\infty}^\infty I_n\Big(\frac{(k_\perp^0)^2}{4\sigma^2}\Big)I_{m-m_\gamma-n}\big((k_\perp^0 w)^2\big)=I_{m-m_\gamma}\Big((k_\perp^{0})^2 w_{eff}^2\Big),\qquad w_{eff}^2:=w^2+\frac{1}{4\sigma^2}.
\end{equation}
Finally, the probability of photoabsorption by the target turns out to be
\begin{equation}\label{P_T1}
	P_T(\mathcal{J}_i\rightarrow\mathcal{J}_f)=1-\exp \Big[-N\frac{4\pi^2 \alpha R_s}{2\mathcal{J}_i+1}  \frac{\sigma_\perp^2e^{-(k_\perp^{0})^2w^2_{eff}}}{\sigma_3(k_3^0)^2} \sum_{j=1}^\infty\sum_{m=-j}^j \big(d^j_{m\lambda_0}(\theta_k^0)\big)^2I_{m-m_\gamma}\Big((k_\perp^{0})^2w^2_{eff}\Big) \sum_{\tau=E,M}|M^\tau_{j}(\varepsilon)|^2\Big].
\end{equation}
If the typical size of the target is small so that all the nuclei are close to the axis of propagation of the twisted photon,
\begin{equation}\label{AssymptTarget}
	k_0 w\ll1, \qquad \frac{k_0}{\sigma} \ll1,
\end{equation}
then
\begin{equation}\label{P_T2}
	P_T(\mathcal{J}_i\rightarrow\mathcal{J}_f)=1-\exp \Big[-N\frac{4\pi^2 \alpha R_s}{2\mathcal{J}_i+1}  \frac{\sigma_\perp^2}{\sigma_3(k_3^0)^2} \sum_{j\geqslant\max(1,|m_\gamma|)}\big(d^j_{m_\gamma\lambda_0}(\theta_k^0)\big)^2\sum_{\tau=E,M}|M^\tau_{j}(\varepsilon)|^2\Big],
\end{equation}
in the leading order. Formulas \eqref{P_T1}, \eqref{P_T2} define the Bouguer law for the absorption of a twisted photon with the energy $\varepsilon$, the opening angle $\theta_k^0$, the projection of the total angular momentum $m_\gamma$, and the helicity $\lambda_0$ .

For large target sizes, when $k_0w\gg1$, one can substitute the modified Bessel function by its asymptotics for large arguments in formula \eqref{P_T1}. Then the sum over the projections $m$ of $\big(d_{m\lambda_0}^j(\theta_k^0)\big)^2$ is equal to one and the probability of photoabsorption becomes
\begin{equation}\label{P_T3}
	P_T(\mathcal{J}_i\rightarrow\mathcal{J}_f)=1-\exp \Big[-\frac{N}{(2\pi)^{\frac12} k_\perp^0 w_{eff}} 
    \frac{4\pi^2 \alpha R_s}{2\mathcal{J}_i+1}
    \frac{\sigma_\perp^2}{\sigma_3(k_3^0)^2} \sum_{j=1}^\infty\sum_{\tau=E,M}|M^\tau_{j}(\varepsilon)|^2\Big].
\end{equation}
The expression in the exponent coincides, up to a factor, with the planewave photoabsorption probability \eqref{prob_photo3_pl}.

\section{Conclusion}

Let us sum up the results. We have obtained the probability of absorption by an atomic nucleus of a photon having a certain projection of the total angular momentum $m_\gamma\in \mathbb{Z}$ onto the propagation axis of a twisted photon. The explicit expressions \eqref{prob_photo1}, \eqref{PhotProbBi}, \eqref{prob_photo3} for the probability of absorption of such a photon by a single nucleus have been derived. It is shown that in the case when the nucleus is placed close to the axis along which the twisted photon propagates, i.e., estimates \eqref{Assympt} are fulfilled, photoabsorption obeys the selection rule, $j\geqslant|m_\gamma|$, where $j$ is the multipolarity of the nucleus transition \cite{MyDiploma,GiantResonance}. In the long-wave approximation, the main contribution to the probability of photoabsorption comes from the transitions with the minimum possible multipolarity $j=|m_\gamma|$. Such a property of absorption of twisted photons by nuclei makes it possible to investigate separately the giant higher-order multipole resonances. We have also obtained the explicit expressions \eqref{P_T1}, \eqref{P_T2}, \eqref{P_T3} for the probability of absorption of the twisted photon by the target consisting of many nuclei. These expressions have the form of the Bouguer law and determine the absorption coefficient of twisted photons by nuclei.

\end{document}